\documentclass{aa}
\usepackage{multirow}
\usepackage{booktabs}
\usepackage[varg]{txfonts}
\usepackage{graphicx}	
\usepackage{amsmath}	
\usepackage{subfigure}
\usepackage{epstopdf}
\usepackage[colorlinks, linkcolor=blue, urlcolor=blue, citecolor=blue]{hyperref}
\def\beq{\begin{equation}}
	\def\enq{\end{equation}}
\def\bea{\begin{eqnarray}}
	\def\ena{\end{eqnarray}}

\def\apjl{ApJL\,}

\begin{document}
	\title{Rotation and dispersion measure evolution of repeating fast radio bursts propagating through a magnetar flare ejecta}
	
	\author{Di Xiao\inst{\ref{inst1}}}
	\institute{Purple Mountain Observatory, Chinese Academy of Sciences, Nanjing 210023, People's Republic of China \label{inst1}\\ \email{dxiao@pmo.ac.cn}}
	
	\date{Received XXX / Accepted XXX}
	\abstract{Rotation measure (RM) and dispersion measure (DM) are characteristic properties of fast radio bursts (FRBs) that contain important information of their source environment. The time evolution of RM and DM is more inclined to be ascribed to local plasma in the host galaxy rather than the intergalactic medium or free electrons in the Milky Way. Recently a sudden drastic RM change was reported for an active repeating FRB 20220529, implying that some kind of mass ejection happened near the source. In this work I suggest that magnetar flare ejecta could play this role and give rise to the significant RM change. I introduce a toy structured ejecta model and calculate the contribution to RM and DM by a typical flare event. I find that this model could reproduce the RM behavior of FRB 20220529 well under reasonable parameters, and similar sudden change is expected as long as this source maintains its activity.}

	\titlerunning{Magnetar flare model for the sudden drastic RM change}
	\authorrunning{D. Xiao}
	\maketitle

	\section{Introduction}
	Fast radio bursts (FRBs) are extremely-bright millisecond pulses discovered in 2007 \citep{Lorimer2007}. Due to the dedicated search for bursts from different radio facilities in the past a few years, more than eight hundred FRB events are discovered and several tens of them are repeating. With the rapidly-accumulating sample of events and bursts, our understanding on this mysterious phenomenon is being refreshed continuously \citep[for recent reviews, see e.g.,][]{XiaoD2021,Petroff2022,ZhangB2023}. The Five-hundred-meter Aperture Spherical Telescope (FAST)  is suitable to perform long-time monitoring of these repeaters. Benefited from its high sensitivity, fine characterization of several repeaters have been achieved, the burst counts of which all reach $\sim$thousand level. 
	
	One of the most interesting findings by FAST is that the dynamic magnetic environment indicated by long-term rotation measure (RM) evolution.  Oscillating RM variation was firstly discovered in FRB 20201124A \citep{XuH2022} and then confirmed in FRB 20190520B even with a sign reversal \citep{Anna-Thomas2023}.  This  oscillation behavior was suggested to originate from a binary system consists of a neutron star (NS) and a massive star \citep{WangFY2022}.  With the orbital motion of the FRB-emitting NS through the magnetized disk or wind of the massive star, significant RM variation is induced then the RM oscillation period is determined by the orbital period \citep{ZhaoZY2023}. This picture remains to be tested since there is no concrete evidence of multiple periods. 
	
	The very recent discovery in FRB 20220529 is more intriguing that a sudden increase of RM was identified. For over 17 months its RM varied slowly between -300 and 300 $\rm rad\,m^{-2}$, however, encountered an abrupt boost to 1976.9 $\rm rad\,m^{-2}$ within two months and recovered to normal level in just 14 days \citep{LiY2025}. There is a long-term DM fluctuation of $\sim20\,\rm pc\,cm^{-3}$ but it is quite stochastic and not closely related to RM variation. Especially, the observed DM variation during the abrupt RM change phase is quite small compared to its stochastic fluctuation. This rapid change of RM indicates that some kind of magnetized plasma appeared in line of sight. The short recovery time suggested a short existence time of this plasma. In the binary scenario, it is likely that the massive star ejected some coronal plasma due to a stellar flare. Alternatively, the magnetar itself can also eject ionized plasma during a flare. Here in this work I focus on this magnetar flare scenario.
	
	The source of FRBs is still on debate but magnetar model is in the leading position since the solid association between an X-ray burst from Galactic magnetar SGR 1935+2154 with FRB 20200428D \citep{CHIME2020b,Bochenek2020,LiCK2021,Mereghetti2020,Ridnaia2021,Tavani2021}. It is well established that magnetar giant flare can eject huge amount of baryons, for instance, an outflow mass of  $\geq10^{24.5}\,\rm g$ for SGR 1806-20 giant flare was confirmed by radio observation \citep{Gelfand2005}. As a scale-down version of giant flare, it is expected that some baryonic plasma was also ejected during normal magnetar flares. Theoretically, baryons in the crust can be ejected via magnetic reconnection induced by starquake or slow untwisting of the internal magnetic field \citep{Thompson1995,Lyutikov2003}. Observationally, a possible indirect evidence is that the persistent radio source of FRB 20121102A and its huge RM is consistent with a flaring magnetar embedded in a wind nebula \citep{Margalit2018}. In this case, sufficient ions are needed since pure pair plasma contributes no net RM. However, an X-ray burst forest was detected before FRB 20200428D \citep{Younes2020,Kaneko2021}, but the RM value of this FRB is similar to those of the normal radio pulses from this magnetar \citep{ZhuWW2023}. This implies that flare ejecta, if exists, does not contribute to the observed RM for this FRB. This point needs to be properly addressed within magnetar flare scenario.
	
	This letter is organized as follows. In Section \ref{sec:model} I introduce a toy model for magnetar flare ejecta. Then in Section \ref{sec:results} I present the theoretical predictions of the model and discuss parameter dependences. Further, I apply our model to two special events in Section \ref{sec:appli}, both the drastic RM change of  FRB 20220529 and the absence of change of FRB 20200428D are well explained. I finish with discussion and conclusions in Section \ref{sec:discuss}.
	
	\section{Toy model description}
	\label{sec:model}
\begin{figure}
		\begin{center}
			\includegraphics[width=\linewidth]{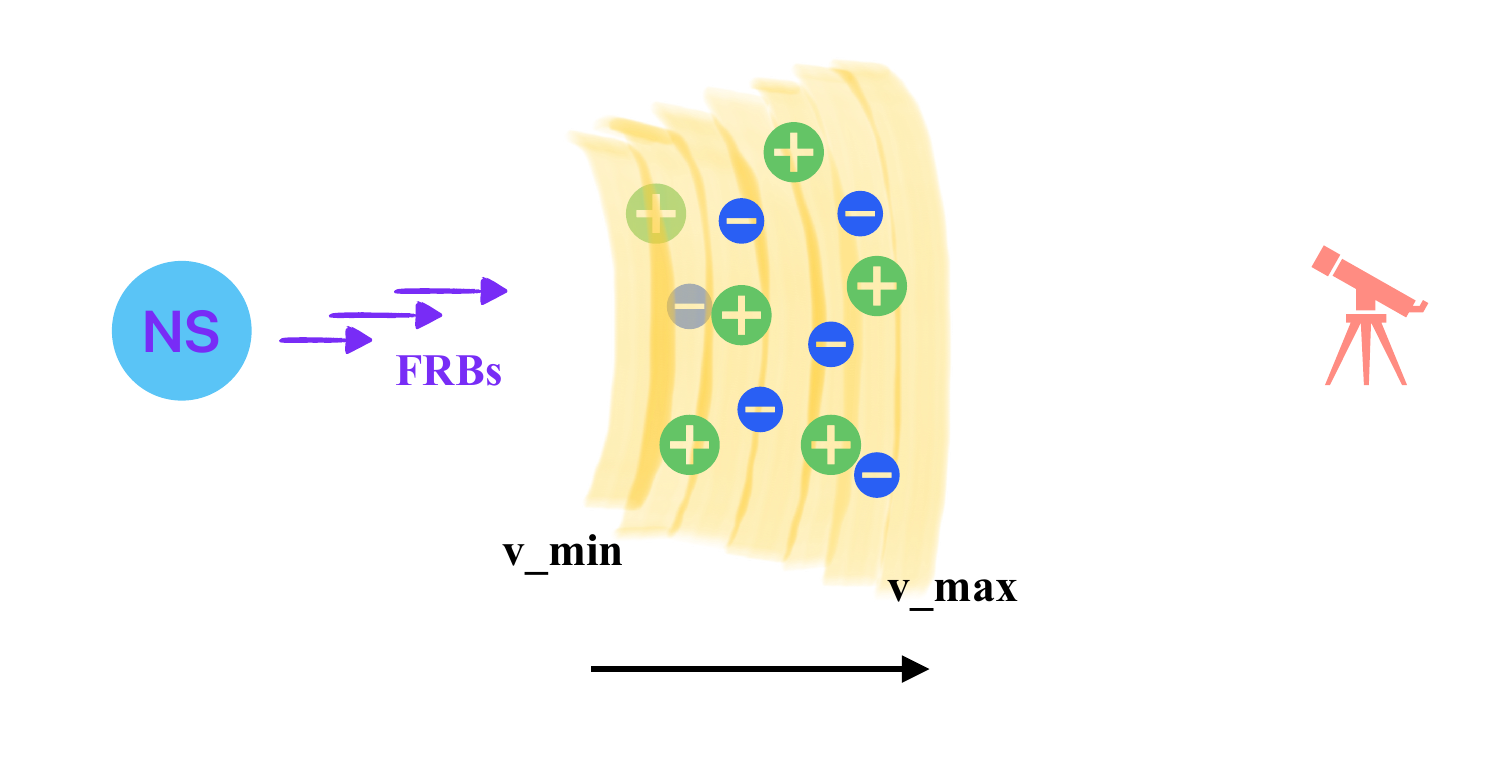}
			\caption{The schematic picture of the toy model. The ejecta is composed of several ion shells with different velocities. The mass distribution is not necessarily uniform in both radial and angular directions.}
			\label {fig:schematic}
		\end{center}
	\end{figure}

	The magnetar can produce a flare and the ejecta may be heavily baryon-loaded. Ejecta matter is assumed to be composed by a bunch of shells with different velocities as shown in Figure  \ref{fig:schematic}. The radial structure of the flare ejecta is assumed to be a universal power-law form, i.e., the mass distribution follows 
	\begin{equation}
		\frac{dM}{dv}\propto v^{s},
	\end{equation}
	with $s\le 0$. The normalization is determined by $\int_{v_{\rm{min}}}^{v_{\rm{max}}}\frac{dM}{dv}dv=M_{\rm{ej}}$ where $v_{\rm{min}},\,v_{\rm{max}}$ are the shell velocities at the inner and outer boundary respectively. The angular structure of ejecta is somewhat uncertain. In general we can introduce the latitudinal and longitudinal angles $\theta_{\rm ej},\,\varphi_{\rm ej}$ to describe. If we assume a homologous expansion $r=vt$, the volume of a shell is 
	\begin{equation}
		dV=r \cos \theta d\theta \cdot rd\varphi \cdot tdv.
	\end{equation}
	Introducing the shell density $\rho$, we have
	\begin{equation}
		\int_{0}^{\varphi_{\rm{ej}}} \int_{v_{\rm{min}}}^{v_{\rm{max}}} \int_{0}^{\theta_{\rm{ej}}} 2\rho \cos \theta v^{2} t^{3} d\varphi dv d\theta = M_{\rm{ej}}.
	\end{equation}
	
	The above expression applies to the most general case. To be more specific, we can made further assumptions on the angular structure. Similar to the structures of gamma-ray burst jet or kilonova ejecta \citep[e.g.,][]{HuangZQ2018}, here I assume a simple power-law mass profile in the latitudinal direction, 
	
	\begin{equation}
		\frac{dm}{\cos \theta d\theta}\propto
		\left\{
		\begin{array}{ll}
			1, & {\rm if}\,\,\theta\leq \theta_{c}, \\
			(\frac{\theta}{\theta_{c}})^{k}, & {\rm if}\,\,\theta>\theta_{c},
		\end{array}
		\right.
	\end{equation}
	where $\theta_{c}$ represents the central core, $dm$ is the mass along a certain longitudinal direction. For simplicity I assume the shell mass is distributed uniformly in longitudinal direction, hence the normalization is calculated by $\int_{0}^{\theta_{\rm{ej}}} 2\frac{dm}{\cos \theta d\theta}d\theta=\frac{dM}{dv}/\varphi_{\rm{ej}}$. Under these simplifications, we can easily write down the shell density,
	\begin{equation}
		\rho=\frac{dm}{\cos \theta d\theta(v^{2}t^{3})}.
	\end{equation}
	Assuming a initially-ionized pure hydrogen ejecta with a mean molecular weight $\mu_m\sim1$ after the magnetar flare,  we get the initial free electron number density as $n_{e,0}=\rho/(\mu_mm_p)$ .  It is reasonable to assume full ionization at the beginning since thermal X-ray emission from the magnetar surface suggests a typical plasma temperature of order $kT_0\sim\rm keV$ \citep{Kaspi2017}, which is much greater than the ionization energy of hydrogen $I_{\rm H}\simeq13.6\,\rm eV$. However, the shells suffer adiabatic cooling while moving outwards, and the temperature decreases with radius as $T\propto r^{-2/3}$. Once it drops close to $I_{\rm H}$, recombination is nonnegligible. The recombination coefficient is given by \citep{Burbidge1963}
	\beq
	\alpha_{\rm R}(T)=2\left(\frac{32\pi}{3^{3/2}}\right)\alpha^3a_0^2\left(\frac{2kT}{\pi m_e}\right)^{1/2}Y\phi(Y)\bar{g}_{\rm R}
	\enq
	where $Y\equiv I_{\rm H}/kT$ is dimensionless, $\alpha$ is fine structure constant, $a_0$ is Bohr radius and $\bar{g}_{\rm R}\sim0.9$ is recombination Gaunt factor. The approximated expression of $\phi(Y)$ is \citep{Tucker1966}
	\begin{equation}
		\phi(Y)\simeq
		\left\{
		\begin{array}{ll}
			0.5[1.735+\ln Y+(6Y)^{-1}], & {\rm if}\,\,Y\geq1, \\
			Y(-0.298-1.202\ln Y),  & {\rm if}\,\,Y\ll 1.
		\end{array}
		\right.
	\end{equation}
	Due to recombination, the decrement rate of number density for electrons and ions is 
	\beq
	-\frac{dn_e}{dt}=-\frac{dn_Z}{dt}=\alpha_{\rm R}n_en_Z.
	\label{eq:ionfrac}
	\enq
	For pure hydrogen ejecta $Z=1$, proton number density $n_Z=n_e$ is achieved at any time. We need to solve the above equation to obtain the number density of free electrons since only they are relevant to DM and RM. The dependence of contributed DM and RM by the flare ejecta with time and viewing angle is 
	\begin{eqnarray}
		\label{eq:evo}
		{\rm DM}&=&\int_{v_{\min}}^{v_{\max}}\frac{n_et\cos{\theta}}{1+z}dv,\nonumber\\
		{\rm RM}&=&\frac{e^3}{2\pi m_e^2c^4}\int_{v_{\min}}^{v_{\max}}\frac{n_eB_{\parallel}t\cos{\theta}}{(1+z)^2}dv,
	\end{eqnarray}
	where $z$ is the redshift of cosmological FRBs. The parallel magnetic field strength is somewhat uncertain. Within the light cylinder radius $R_{\rm LC}=cP/2\pi$, a dipole configuration is usually adopted. Generally, the field configuration outside the light cylinder can be obtained by solving an axially symmetric force-free pulsar equation for the magnetic flux \citep{Mestel1973,Michel1973a,Okamoto1974}. Two of the analytic solutions found were ``split monopole'' \citep{Michel1973b} or ``inclined split monopole'' \citep{Bogovalov1999}, in which the poloidal field strength decays as $B_p\propto r^{-2}$ and the toroidal field decays as $B_\varphi\propto r^{-1}$. Further, 3D numerical time-dependent calculations for the force-free magnetosphere of an oblique rotator were performed and the ``universal solution'' was found \citep{Spitkovsky2006, Kalapotharakos2009}. Although it differs from Michel-Bogovalov monopole for sufficiently large inclination angles, the poloidal field averaged over the angle $\varphi$ still had the form of $\left<B_p\right>\propto r^{-2}$. This universal solution was generally reproduced via particle-in-cell simulations \citep{Philippov2015}. In our scenario, FRBs and flare ejecta are both produced by the central magnetar and then propagate radially. Therefore, only the poloidal field component can contribute to the observed RM. Therefore I consider a generic description with a smooth transition at $R_{\rm LC}$, 
	\begin{equation}
		B_{\parallel}(r)\simeq
		\left\{
		\begin{array}{ll}
			B_0(r/R_0)^{-3}, & {\rm if}\,\,r<R_{\rm LC}, \\
			B_0(R_{\rm LC}/R_0)^{-3}(r/R_{\rm LC})^{n}, & {\rm if}\,\,r>R_{\rm LC},
		\end{array}
		\right.
		\label{eq:Bparall}
	\end{equation}
	with a fixed $n=-2$. Note that Eq.(\ref{eq:evo}) is only valid for cold plasma \citep{Yang2023}. However, the ejected plasma from the magnetosphere is initially hot and ultra-relativistic. Outside the magnetosphere, there is no parallel electric field to accelerate charged particles.  Adopting canonical parameters of $B_p=10^{15}\,\rm G$ and $P=1\,\rm s$ for a magnetar, the field strength at $R_{\rm LC}$ is $B_{\rm LC}=9.2\times10^3\,\rm G$. Then these relativistic electrons can cool efficiently due to synchro-curvature radiation.  We can evaluate the synchrotron cooling timescale as 
	\begin{equation}
		t_{\rm syn}=\frac{6\pi m_ec}{\sigma_TB^2\gamma_e}=3.08\times10^{-2}B_4^{-2}\gamma_{e,2}^{-1}\,\rm s,
	\end{equation}
	therefore on the $\sim$day timescale of RM evolution like FRB 20220529, the relativistic electrons have cooled down,  then Eq.(\ref{eq:evo}) is justified.
	
	The other issue is that FRBs may be absorbed by the ejecta. The free-free absorption opacity $\tau_{\rm ff}$ is
	\beq
	{\rm \tau_{\rm ff}}=\int_{v_{\min}}^{v_{\max}}{n_e^2\alpha_{\rm ff}t \cos{\theta}dv},
	\enq
	where the free-free absorption coefficient and Gaunt factor are functions of temperature \citep{Yang2017}
	\bea
	\alpha_{\rm ff}&=&\frac{4}{3}\left(\frac{2\pi}{3}\right)^{1/2}\frac{Z^2e^6n_en_Z\bar{g}_{\rm ff}}{cm_e^{3/2}(k_BT)^{3/2}\nu^2},\nonumber\\
	\bar{g}_{\rm ff}&=&\frac{\sqrt{3}}{\pi}\left[\ln\left(\frac{(2k_BT)^{3/2}}{\pi e^2m_e^{1/2}\nu}\right)-\frac{5}{2}\gamma\right],
	\ena
	where $\nu$ is the frequency of FRBs and $\gamma=0.577$ is Euler's constant.  With all the ingredients ready, we can calculate the DM and RM as FRBs propagate through flare ejecta.
	
	\section{Results and parameter dependences}
	\label{sec:results}
\begin{figure}
		\begin{center}
			\includegraphics[width=0.95\linewidth]{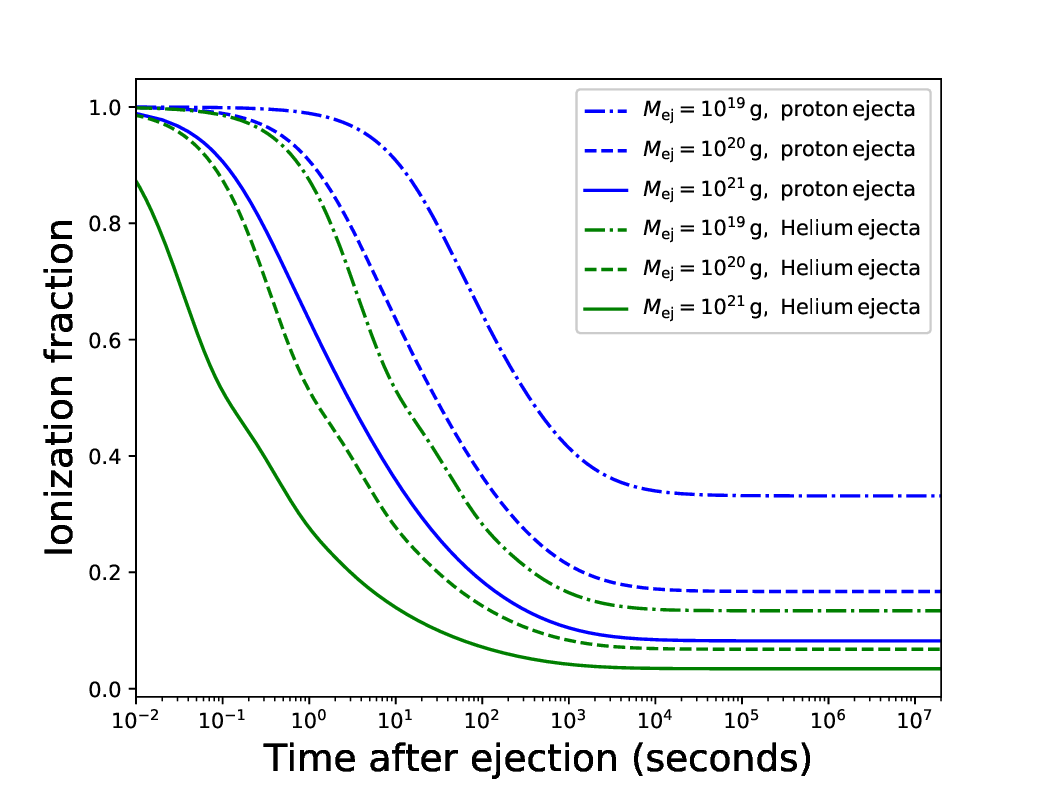}
			\caption{The time evolution of the ionization fraction of the ejecta. Different line styles represent cases of different ejecta mass, while different colors represent different composition. Recombination effect is important and generally the heavier the ejecta, the lower the ionization fraction. For pure helium ejecta, the ionization fractions are much lower than those of hydrogen ejecta, giving rise to smaller DM and RM contributions.}
			\label {fig:ionfrac}
		\end{center}
	\end{figure}

\begin{figure}
		\begin{center}
			\includegraphics[width=0.95\linewidth]{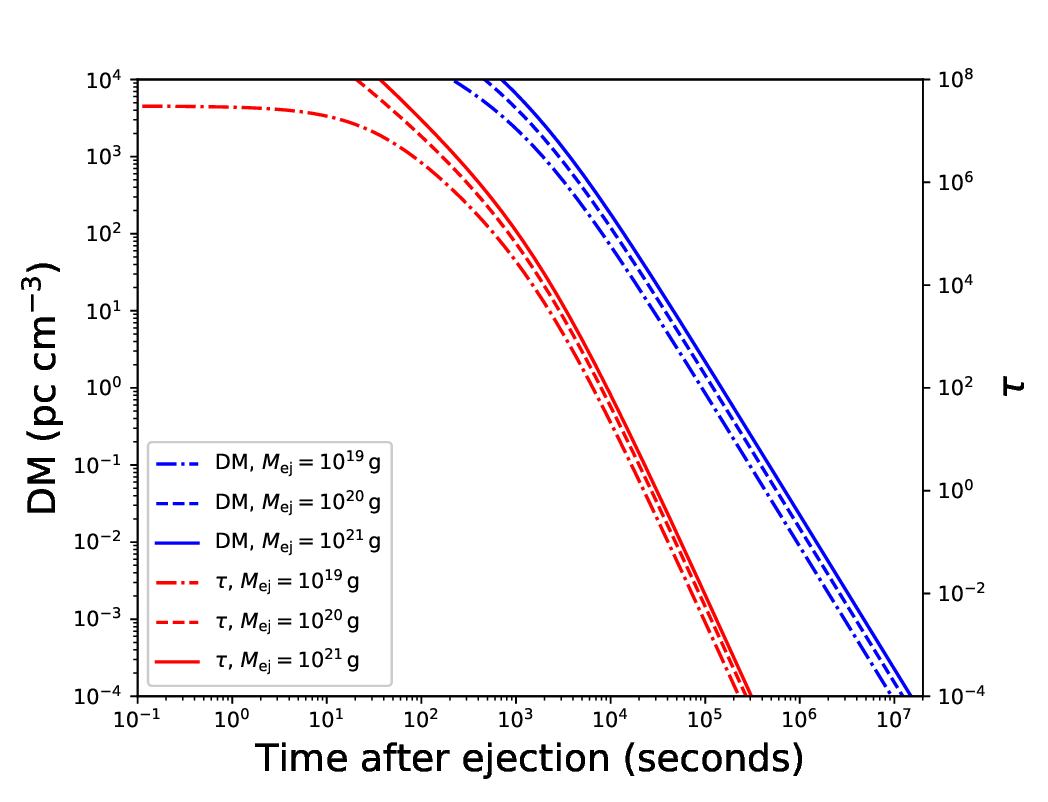}
			\caption{The time evolution of the DM and $\tau_{\rm ff}$ contributed by a typical hydrogen ejecta.}
			\label {fig:DM_fid}
		\end{center}
	\end{figure}

\begin{figure}
		\begin{center}
			\includegraphics[width=0.95\linewidth]{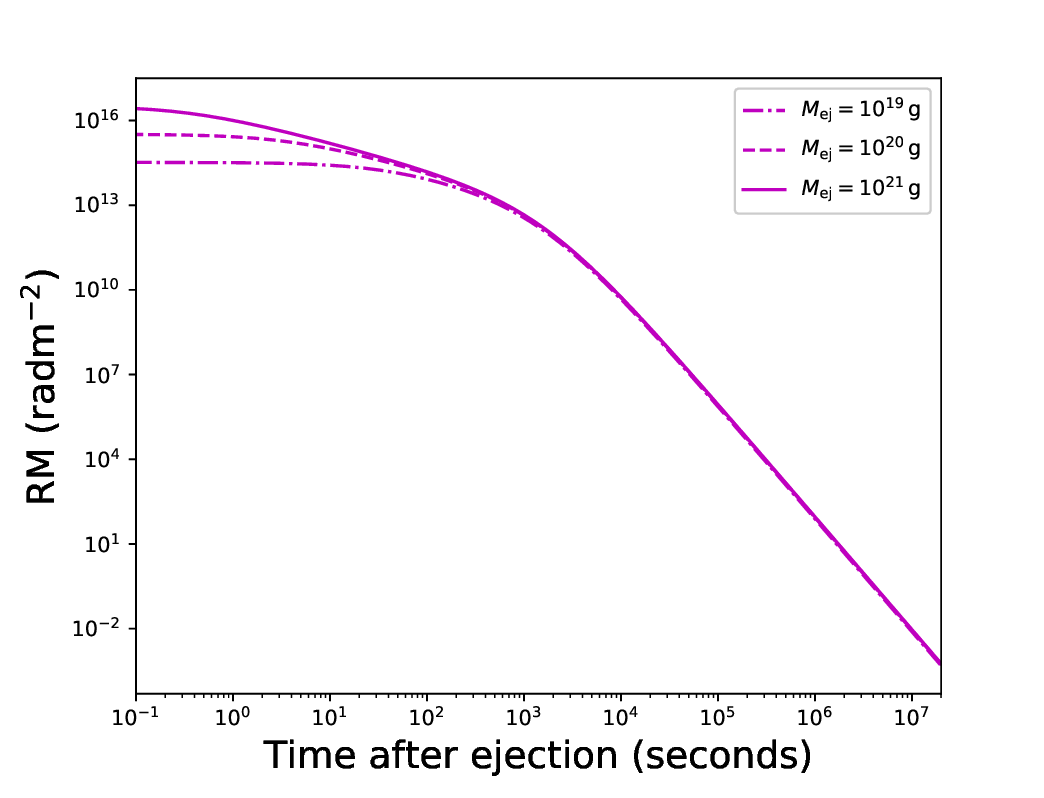}
			\caption{The time evolution of the RM contributed by a typical ejecta.}
			\label {fig:RM_fid}
		\end{center}
	\end{figure}

	Here I show the calculated DM and RM for the most general cases. Typical parameters are assumed as $B_0=10^{15}\,\rm G$, $v_{\max}=0.1c,\, v_{\min}=0.0001c,\,s=-2$. I assume the simplest spherically symmetric angular profile, i.e., $\theta_{\rm ej}=\pi/2,\,\varphi_{\rm ej}=\pi,\,k=0$, therefore the viewing angle is $\theta=0$. First  we can solve Eq.(\ref{eq:ionfrac}) to get the ionization fraction evolution, which is shown in Fig. \ref{fig:ionfrac}. Obviously it depends on the total ejecta mass and I compare three cases of different $M_{\rm ej}$ values. Right after the ejection from magnetosphere, recombination start to play an important role in less than 10s and ionization fraction reaches a constant value after $\sim1000$s. The heavier the ejecta, the lower the fraction in equilibrium. The ejecta composition can also influence the ionization fraction and then the number density of free electrons. For other hydrogen-like ions with charge number $Z$,  the mean molecular weight is different and the expression of $Y$ in Eq.(6) should multiply $Z^2$. In Fig. \ref{fig:ionfrac} I also show the cases of pure helium ejecta for comparison. The ionization fractions of helium ejecta are much lower, therefore the contributed DM and RM are smaller in proportion. Further we can obtain the time evolution of DM, RM and $\tau_{\rm ff}$ that are shown in Fig. \ref{fig:DM_fid} and Fig. \ref{fig:RM_fid}. We can see that these quantities all increase with $M_{\rm ej}$, but only to a limited extent due to the decrease of ionization fraction. The characteristic timescale for the ejecta to be transparent to free-free absorption is a few thousand seconds. At this moment, the DM contributed by the ejecta is only a few units, therefore it is hard to identify a flare event from DM variations of a repeating FRB. However, the contribution to RM by the ejecta can be very prominent. For our fiducial case in Fig. \ref{fig:RM_fid},  the RM contribution is still high ($\geq 10^3\,\rm rad\,m^{-2}$) after $\sim10$ days, which is much easier to identify in observations. I show in the next section that the sudden dramatic RM change of FRB 20220529 is probably a signature of flare ejecta. 
	
	
\begin{figure}
		\begin{center}
			\includegraphics[width=0.95\linewidth]{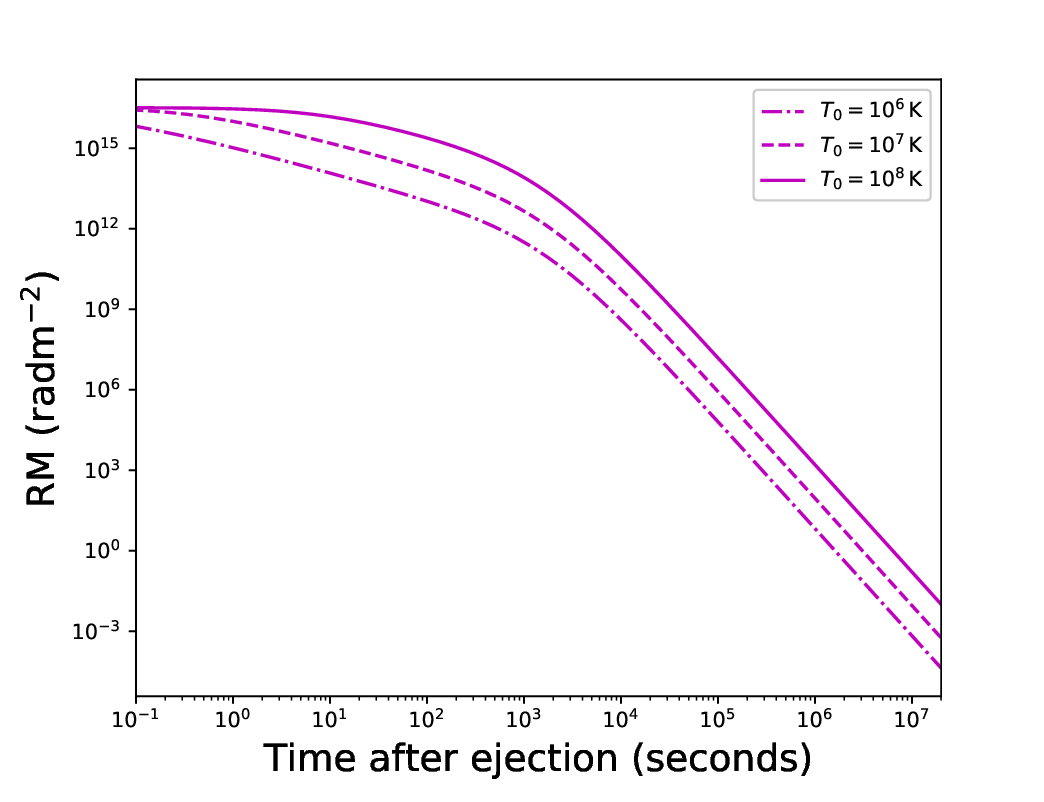}
			\caption{Similar to Fig. \ref{fig:RM_fid} but fixing $M_{\rm ej}=10^{21}\,\rm g$ and letting $T_0$ vary.}
			\label {fig:diffT}
		\end{center}
	\end{figure}

	Other than $M_{\rm ej}$, there are many more parameters that can have a impact on the RM evolution. The influence of $B_0$ is quite straightforward. The spin period determines the light cylinder radius, therefore will also influence RM according to Eq. (\ref{eq:Bparall}). The choice of $v_{\max},\,v_{\min}$ influence the decay timescale of RM and DM. The time needed for the contribution becomes negligible is shorter for ejecta moving out with higher speed. Furthermore, the initial temperature of the ejecta will influence the recombination rate, thus indirectly influence the RM values. This is shown in  Fig. \ref{fig:diffT} that the higher initial temperature, the more abundant free electrons and the greater RM.
	
	\section{Applications to two special FRB events}
	\label{sec:appli}
\begin{figure}
		\centering
		\subfigbottomskip=2pt
		\subfigcapskip=-1pt
		\subfigure[]{
			\includegraphics[width=0.95\linewidth]{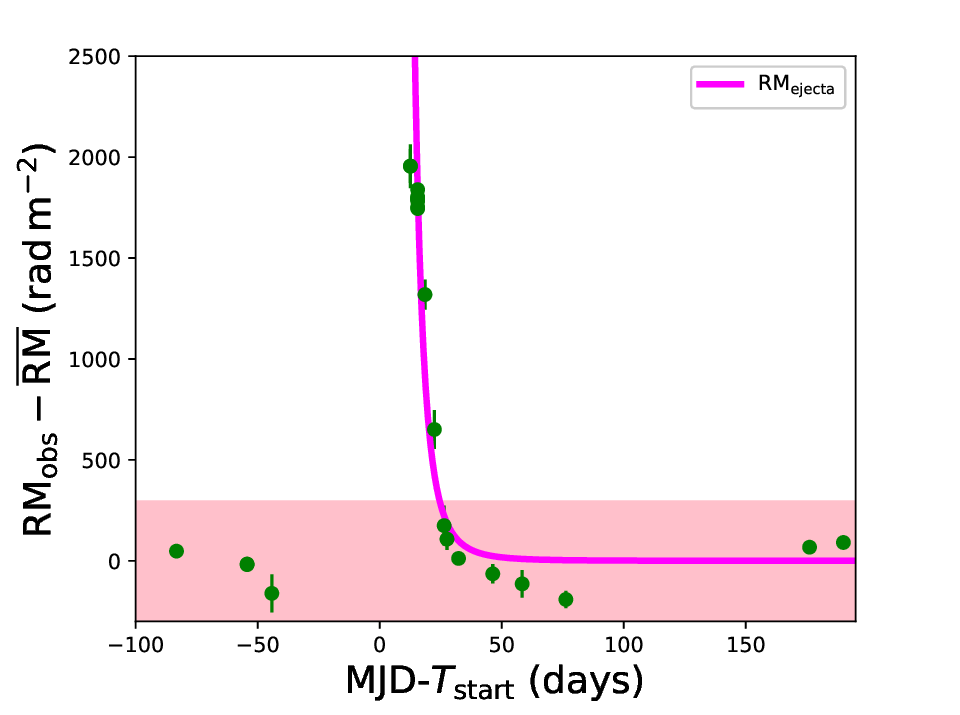}
			\label{fig:modelfit}
		}
	\vskip -3pt
		\subfigure[]{
			\includegraphics[width=0.95\linewidth]{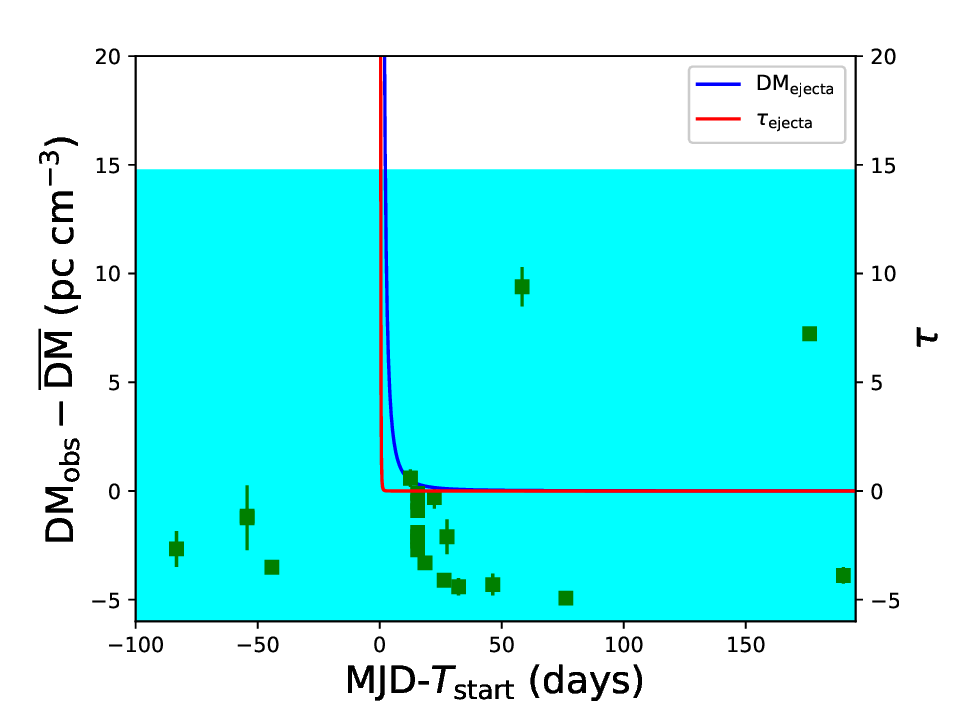}
			\label{fig:append}
		}
		\caption{Fitting results of FRB 20220529 by the flare scenario. Panel(a): Model fitting of the RM behavior. A baseline value $\overline{\rm RM}=21.2\,\rm rad\,m^{-2}$ has been subtracted. The shaded pink region shows the RM variation range outside the abrupt RM change period. Panel(b): Time evolution of DM and $\tau_{\rm ff}$ under the best fitting parameters in Table \ref{tb:bestfit}. The DM contribution by the ejecta is very small compared to the stochastic fluctuation shown by the shaded cyan area.}
		\label{fig:total}
	\end{figure}

	In this section I will discuss the application of our model to two special FRBs.  The rapid change of RM is a natural outcome of matter ejection according to above section. I assume the magnetar flare occurs at $T_{\rm start}$ (in MJD). Taking $(M_{\rm ej},\, B_0,\, P_0, \, T_{\rm start} )$ as parameters and fixing $s=-1,\, v_{\min}=0.0001c,\, v_{\max}=0.1c,\, T_0=10^8\,\rm K$, I use a Malkov-chain-Monte-Carlo approach and the RM data can be well fitted, as shown in Fig. \ref{fig:modelfit}. At the same time $\tau_{\rm ff}$ and DM contributed by the ejecta are both negligible (see Fig. \ref{fig:append}). Outside the period of the abrupt RM change phase, there are stochastic fluctuations of RM and DM with mean values of $\overline{\rm RM}=21.2\,\rm rad\,m^{-2}$, $\overline{\rm DM}=249.4\,\rm pc\,cm^{-3}$, which are adopted as their baseline values respectively. The pink and cyan area in Figure \ref{fig:total} show the long-term variation ranges with respect to the baseline values. The best-fitting parameters are listed in Table \ref{tb:bestfit} and the relevant corner plot is shown in Fig. \ref{fig:corner}.  All parameters are in reasonable range, therefore, the magnetar flare scenario can explain this special RM behavior. 
	
	\begin{table}
		\centering
		\caption{The best-fitting values using MCMC method.}
		\label{tb:bestfit}
		\begin{tabular}{ccc}
			\toprule Parameter &Allowed range & Best-fitting value \\
			\midrule$\log M_{\rm ej}$ &[18.0, 24.0]& $ 21.96_{-1.17}^{+1.33}$ \\
			$\log B_0$ &[14.0, 16.0] & $ 15.69_{-0.32}^{+0.22}$ \\
			$\log P_0$ &[-1.0, 1.0] & $ 0.16_{-0.28}^{+0.40}$ \\
			$T_{\rm start}$ (MJD) &[60280, 60292]& $ 60280.004_{-0.0034}^{+0.0075}$ \\
			\bottomrule & &
		\end{tabular}
	\end{table}

\begin{figure}
		\begin{center}
			\includegraphics[width=0.95\linewidth]{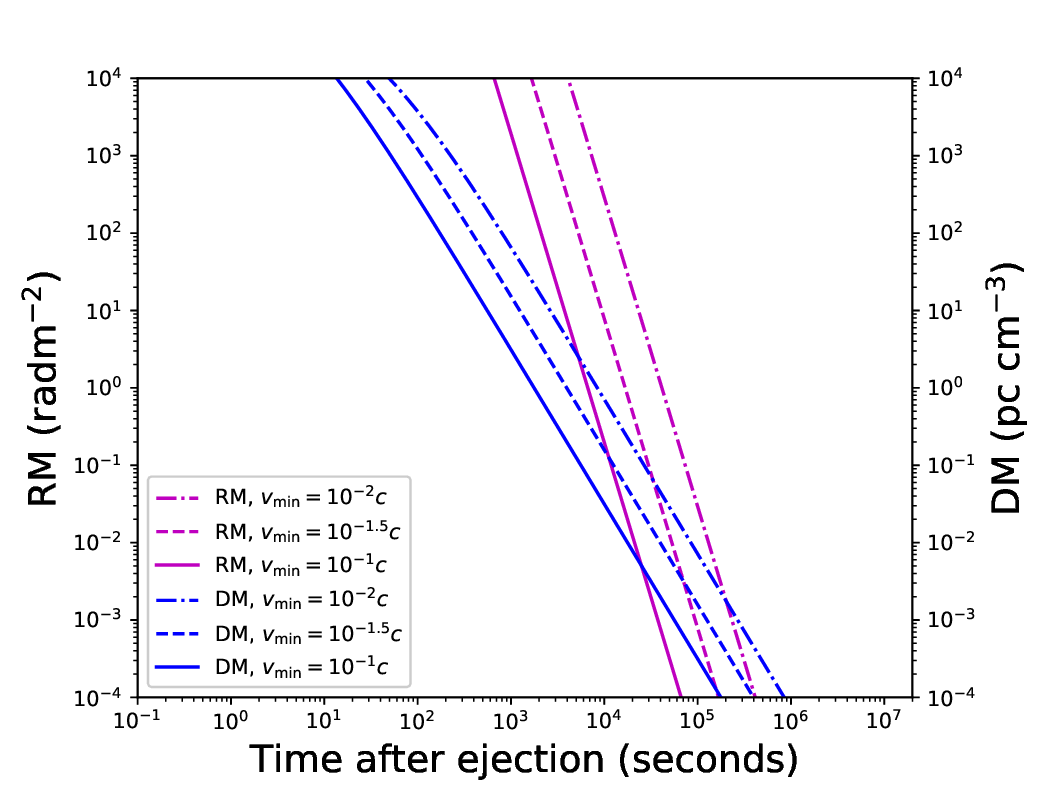}
			\caption{One possible set of parameters explaining the absence of notable RM change for FRB20200428D. Both RM and DM contribution decline to unity rapidly, therefore can hardly be identified in observation.}
			\label {fig:1935}
		\end{center}
	\end{figure}

	Different from FRB 20220529, no significant RM change was observed for the Galactic FRB 20200428D, which should also be explained within this scenario. As I have discussed above, the amplitude and decay timescale of RM depend on several parameters.  For the source magnetar SGR 1935+2154, the spin period $P_0=3.245\,\rm s$ and magnetic field strength $B_0\simeq2.2\times10^{14}\,\rm G$ have been determined from observation \citep{Israel2016}.  Therefore, we can adjust other parameters to reduce the RM contribution. As an example, I fix $M_{\rm ej}=10^{21}\,\rm g$, $T_0=10^6\,\rm K$, $s=-2$, $v_{\max }=0.3c$ and let $v_{\min}$ vary. A relative large $v_{\max }$ is adopted for this magnetar to ensure that the total kinetic energy of flare ejecta matches with its observed X-ray burst energy of $\sim10^{39}\,\rm erg$. The result is shown in Fig. \ref{fig:1935}. Just after $\sim 10^4\,\rm s$, both RM and DM fall below unity. If we do not luckily catch an FRB in such short timescale after the flare, the signal of RM contribution by the ejecta is probably missed in observation. This is probably the case for FRB 20200428D since it is not a very active repeater. There is an X-ray burst accompanying FRB 20200428D that might indicate matter ejection. However,  detailed analysis suggested that the radio burst occurred slightly earlier than X-ray burst, therefore, these ejected shells will have no influence on DM and RM. 
	
	\section{Discussion and conclusions}
	\label{sec:discuss}
	In this work I constructed a simple toy model for the structure of magnetar flare ejecta and calculated its possible contribution to RM and DM values of FRBs.  I find that for canonical magnetar flare parameters, it is possible to identify the RM change from observations. It's more difficult to identify the DM change since after the time of $\tau_{\rm ff}<1$, DM contribution by the ejecta is only moderate $(\lesssim10\,\rm pc\,cm^{-3})$ and decays rapidly. I note that there is a suitable ``time window" to identify these variation behaviors in observation. Below $\sim0.1 $days the ejecta is still opaque due to free-free absorption and above a few tens of days the RM contribution by the ejecta becomes negligible. Assuming that the magnetar flare ejection (X-ray activity) and FRB production are physically unrelated, it is easier for an active repeater to have a burst falling in this time window. In this sense, it is a natural expectation that this RM change was observed for FRB 20220529 but not for FRB 20200428D due to a difference in source activity. Moreover, the relevant parameters of FRB 20220529 source magnetar may be more preferred to give rise to RM change. Therefore, as long as FRB 20220529 maintains its activity, more similar sudden RM changes are guaranteed to be discovered by long-time monitoring of this source.
	
	In this work I adopt the simplest case that the ejecta is spherically symmetric and the viewing angle $\theta=0$. In principle the angular structure can be more complicated. Due to the limited number of bursts with RM  measurements in flare stage, the angular structure cannot be constrained well and model fitting by the simple case is good enough. If we observe another similar case with adequate datapoints characterizing the RM evolution in great detail in the future, probing the structure of ejecta will be realizable. One critical advantage of our model is that the magnetar itself can get it done without a companion, therefore the occurrence rate of similar RM change is higher than that of binary model. The probability of a companion ejecting coronal mass at right time, right direction is very low.
	
	Since there is a certain chance to observe sudden RM change contributed by the flare ejecta, I encourage to search for FRBs following X-ray bursts of magnetars within a time lag of $\sim 10^5\,\rm s$. SGR 1935+2154 has a relatively high activity in X-rays but it only emit FRBs occasionally.  The redshift of FRB 20220529 source is a bit high ($z=0.1839$), thus the X-ray flux is  too low to be detected by current X-ray satellites assuming typical burst luminosity of $\sim10^{41}\,\rm erg$.  Therefore we can hardly forecast when a similar sudden RM change will occur without X-ray notice. We hope for a nearby golden event of an active repeating FRB associated with enormous X-ray activity in the future, which will test our model in detail and help break the degeneracy of parameters.

	\begin{acknowledgements}
		
		I am grateful to an anonymous referee for helpful comments. This work is supported by the National Natural Science Foundation of China (Grant Nos. 12373052, 12321003, 12393813) and the National SKA Program of China (2022SKA0130100).
		
	\end{acknowledgements}
	
	\bibliographystyle{aa}
	\bibliography{FRBlatest}

\begin{appendix}
\section{Parameter corner plot of the MCMC results.}

	\begin{figure}[h]
	\begin{center}
		\includegraphics[width=\linewidth]{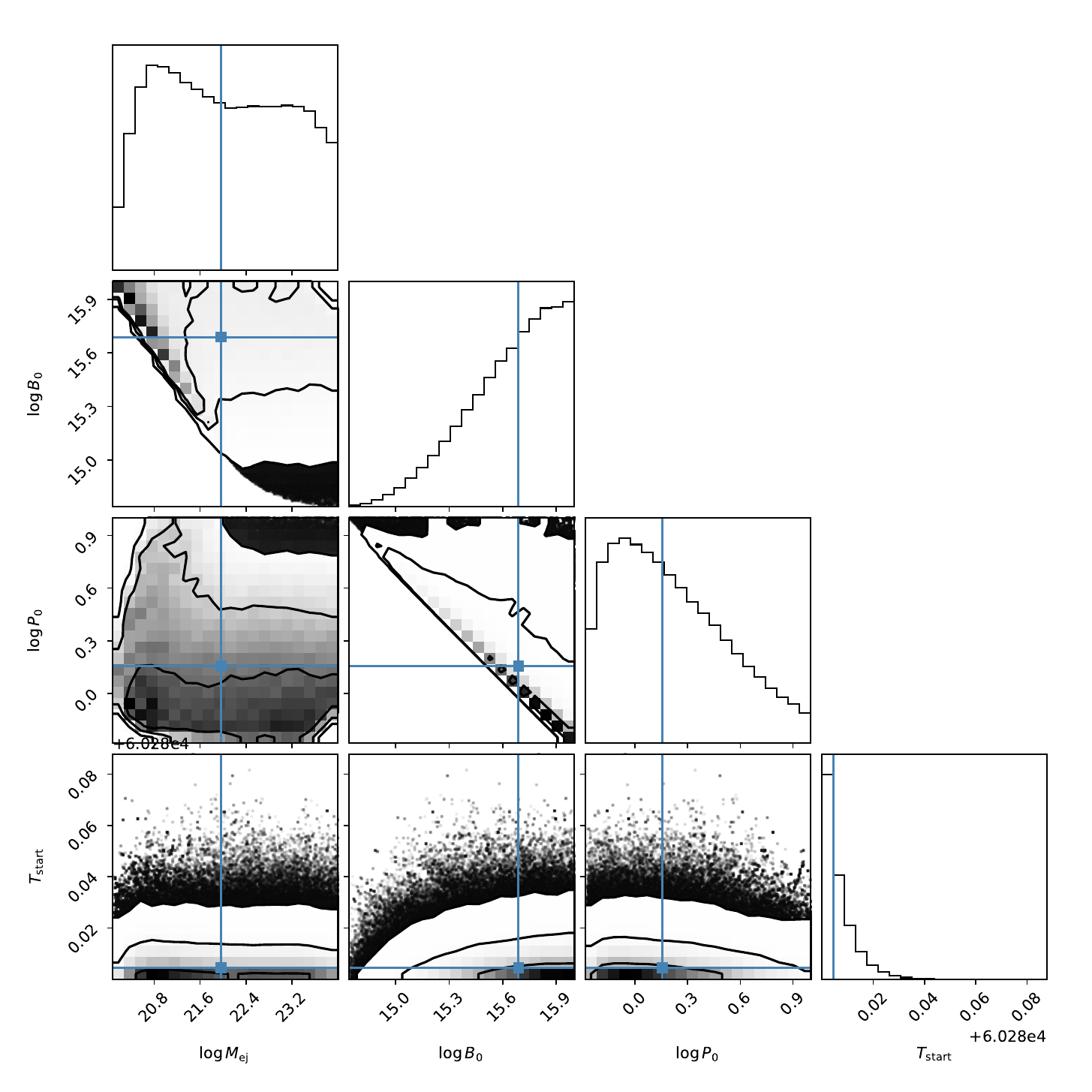}
		\caption{Parameter corner plot of the MCMC results. The contours are 1$\sigma$, 2$\sigma$ and 3$\sigma$ uncertainties. }
		\label {fig:corner}
	\end{center}
	\end{figure}
\end{appendix}
	
\end{document}